\documentclass[runningheads]{llncs}
\usepackage{graphicx}
\usepackage{todonotes}
\usepackage{listings}
\usepackage{amsmath,amssymb,amsfonts}
\usepackage{algorithmic}
\usepackage{graphicx}
\usepackage{textcomp}
\usepackage{svg}
\usepackage{moresize}
\usepackage{times}
\usepackage{color}
\usepackage[multiple]{footmisc}
\usepackage{xcolor}
\def\BibTeX{{\rm B\kern-.05em{\sc i\kern-.025em b}\kern-.08em
    T\kern-.1667em\lower.7ex\hbox{E}\kern-.125emX}}
\definecolor{darkgreen}{rgb}{0.0,0.4,0.0} 
\definecolor{darkred}{rgb}{0.6,0.1,0.1}
\definecolor{lightgray}{gray}{.98}
\definecolor{medgray}{gray}{.70}
\definecolor{darkgray}{gray}{.40}
\definecolor{lightviolet}{rgb}{0.7,0,0.7} 
\definecolor{lightlightviolet}{rgb}{1.0,0.7,1.0} 
\definecolor{darkviolet}{rgb}{0.5,0.1,0.5}
\definecolor{darkredviolet}{rgb}{0.6,0.1,0.4}
\definecolor{limegreen}{rgb}{0.2,0.7,0.2}
\definecolor{navyblue}{RGB}{0,0,128}
\definecolor{aquamarine}{RGB}{102,205,170}

\definecolor{strictRED}{RGB}{184,0,0}
\definecolor{specificationTURQUOISE}{RGB}{0,128,153}
\definecolor{assumptionGREEN}{RGB}{0,128,0}
\definecolor{interruptBLUE}{RGB}{0,0,128}
\definecolor{committedORCHID}{RGB}{54,22,89}
\definecolor{urgentORCHID}{RGB}{74,28,109}
\definecolor{requestedORCHID}{RGB}{104,34,139}
\definecolor{eventuallyORCHID}{RGB}{154,50,205}

\lstdefinelanguage{SMLX}
{
	basicstyle=\ssmall\ttfamily, % 
	frame=single, %none|leftline|topline|bottomline|lines|single|shadowbox
	framextopmargin=0pt,
	framexbottommargin=0pt,
	framexleftmargin=0pt,
	xleftmargin=16pt,
	xrightmargin=3pt,
	morekeywords=[1]{system, domain, scenario, bind, to, 
		message, non, spontaneous, events, specification, 
		alternative, if, collaboration, role, with, dynamic, 
		bindings, or, and, null, define, as, 
		constraints, import, static, parameter, ranges, var, EInt, 
		controllable},
	morekeywords=[2]{strict},
	morekeywords=[3]{forbidden, violation},
	morekeywords=[4]{interrupt},
	morekeywords=[5]{guarantee},
	morekeywords=[6]{assumption}, %obsolete due to moredelim hack below
	morekeywords=[7]{committed}, %obsolete due to moredelim hack below
	morekeywords=[8]{urgent},
	morekeywords=[9]{requested},
	morekeywords=[10]{eventually},
	keywordstyle=[1]\color{darkviolet}\textbf,
	keywordstyle=[2]\color{strictRED}\textit,
	keywordstyle=[3]\color{strictRED}\textit,
	keywordstyle=[4]\color{interruptBLUE}\textit,
	keywordstyle=[5]\color{specificationTURQUOISE}\textbf,
	keywordstyle=[6]\color{assumptionGREEN}\textbf,
	keywordstyle=[7]\color{committedORCHID}\textit,
	keywordstyle=[8]\color{urgentORCHID}\textit,
	keywordstyle=[9]\color{requestedORCHID}\textit,
	keywordstyle=[10]\color{eventuallyORCHID}\textit,
	sensitive=false,
	morecomment=[l][\color{darkgreen}\textit]{//},
	morecomment=[s][\color{darkgreen}\textit]{/*}{*/}, 
	morestring=[b][\color{blue}]",
	tabsize=1,
	moredelim = [s][\color{specificationTURQUOISE}\textbf]{guarantee}{scenario},
	moredelim = [s][\color{assumptionGREEN}\textbf]{assumption}{scenario},
	backgroundcolor=\color{lightgray}
}

\lstdefinestyle{SMLXStyle} {language=SMLX}

\lstdefinelanguage{SMLConfig}
{
	basicstyle=\ssmall\ttfamily,
	frame=single, %none|leftline|topline|bottomline|lines|single|shadowbox
	framextopmargin=0pt,
	framexbottommargin=0pt,
	framexleftmargin=0pt,
	xleftmargin=16pt,
	xrightmargin=3pt,
	morekeywords=[1]{symbolic, import, configure, specification, use, 
		instancemodel, symbolic, parameters, attributes, symbolic, state, matching, 
		off, under, approximation, on, rolebindings, collaboration, object, plays, 
		role, role1},
	keywordstyle=[1]\color{darkviolet}\textbf,
	sensitive=false,
	morecomment=[l][\color{darkgreen}\textit]{//},
	morecomment=[s][\color{darkgreen}\textit]{/*}{*/}, 
	morestring=[b][\color{navyblue}\textit]",
	stringstyle=\color{navyblue},
	tabsize=1,
	backgroundcolor=\color{lightgray}
}

\lstdefinestyle{SMLConfigStyle} {language=SMLConfig}
\lstdefinelanguage{Java}
{
	basicstyle=\ssmall\ttfamily,
	frame=single, %none|leftline|topline|bottomline|lines|single|shadowbox
	framextopmargin=0pt,
	framexbottommargin=0pt,
	framexleftmargin=0pt,
	xleftmargin=16pt,
	xrightmargin=3pt,
	morekeywords=[1]{public, private, class, extends, protected, void,
		new, throws, null, if, else},
	morekeywords=[2]{STRICT},
	morekeywords=[3]{@Override},
	morekeywords=[4]{car, oc, cp}, % local fields
	keywordstyle=[1]\color{darkviolet}\textbf,
	keywordstyle=[2]\color{javablue}\textbf,
	keywordstyle=[3]\color{darkgray},
	keywordstyle=[4]\color{navyblue},
	sensitive=false,
	morecomment=[l][\color{javagreen}\textit]{//},
	morecomment=[s][\color{javagreen}\textit]{/*}{*/}, 
	morestring=[b][\color{javablue}\textit]",
	stringstyle=\color{navyblue},
	tabsize=1,
	backgroundcolor=\color{lightgray}
}

\definecolor{javared}{rgb}{0.6,0,0} % for strings
\definecolor{javablue}{rgb}{0,0,0.9} % for strings
\definecolor{javagreen}{rgb}{0.25,0.5,0.35} % comments
\definecolor{javapurple}{rgb}{0.5,0,0.35} % keywords
\definecolor{javadocblue}{rgb}{0.25,0.35,0.75} % javadoc

\lstdefinestyle{JavaStyle} {language=Java}

\lstdefinelanguage{Kotlin}{
	basicstyle=\ssmall\ttfamily,
	frame=single, %none|leftline|topline|bottomline|lines|single|shadowbox
	framextopmargin=0pt,
	framexbottommargin=0pt,
	framexleftmargin=0pt,
	xleftmargin=16pt,
	xrightmargin=3pt,
	comment=[l]{//},
	commentstyle={\color{darkgray}\ttfamily},
	emph={delegate, filter, first, firstOrNull, forEach, lazy, map, mapNotNull, println, return@},
	emphstyle={\color{darkviolet}},
	identifierstyle=\color{black},
	numberstyle=\color{darkgreen},
	keywords=[1]{ abstract, actual, as, as?, break, by, class, companion, continue, data, do, dynamic, else, enum, expect, false, final, for, fun, get, if, import, in, interface, internal, is, null, object, override, package, private, public, return, set, super, suspend, this, throw, true, try, typealias, val, var, vararg, when, where, while},
	keywordstyle=[1]{\color{javablue}\bfseries},
	keywords=[2]{@Deprecated, @JvmField, @JvmName, @JvmOverloads, @JvmStatic, @JvmSynthetic, @Test, Array, Byte, Double, Float, Int, Integer, Iterable, Long, Short, String},
	keywordstyle=[2]{\color{javablue}},	
	keywords=[3]{interruptingEvents, forbiddenEvents, it}, %immutable properties
	keywordstyle=[3]{\color{darkviolet}\bfseries},
	keywords=[4]{scenario, cycleScenario, runTest, Given, When, Then, And, But}, % 
	keywordstyle=[4]{\textit},
	keywords=[5]{coolantTemp, deratingFactor}, % mutable properties
	keywordstyle=[5]{\color{darkviolet}\bfseries\underbar},
	keywords=[6]{currentTemp}, % mutable properties
	keywordstyle=[6]{\underbar},
	morecomment=[s]{/*}{*/},
	morecomment=[s][\color{black}]{`}{`},
	morestring=[b]",
	morestring=[s]{"""*}{*"""},
	sensitive=true,
	stringstyle={\color{javagreen}\ttfamily},
}

\lstdefinestyle{KotlinStyle} {language=Kotlin}

\lstdefinelanguage{Gherkin}{
	basicstyle=\ssmall\ttfamily,
	frame=single, %none|leftline|topline|bottomline|lines|single|shadowbox
	framextopmargin=0pt,
	framexbottommargin=0pt,
	framexleftmargin=0pt,
	xleftmargin=16pt,
	xrightmargin=3pt,
	comment=[l]{//},
	commentstyle={\color{darkgray}\ttfamily},
	emph={@SoSLevel, @RpsSystem },
	emphstyle={\color{limegreen}},
	identifierstyle=\color{black},
	keywords={Given, When, Then, And},
	keywordstyle={\color{violet}\ttfamily},
	morecomment=[s]{/*}{*/},
	morestring=[b]",
	morestring=[s]{"""*}{*"""},
	ndkeywords={Scenario, Example, Feature},
	ndkeywordstyle={\color{darkviolet}\bfseries},
	sensitive=true,
	stringstyle={\color{javagreen}\ttfamily},
}

\lstdefinestyle{GherkinStyle} {language=Gherkin}

\makeatletter
\lstdefinestyle{mystyle}{
	basicstyle=%
	\ttfamily
	\lst@ifdisplaystyle\footnotesize\fi
}

\makeatletter
\newcommand\applyCurrentFontsize
{%
  \let\f@sizeS@ved\f@size%
  \let\f@baselineskipS@ved\f@baselineskip%
  \let\basicstyleS@ved\lst@basicstyle%
  \renewcommand\lst@basicstyle%
  {%
      \basicstyleS@ved%
      \fontsize{\f@sizeS@ved}{\f@baselineskipS@ved}%
      \selectfont%
  }%
}
\makeatother

\newcommand\scaledlstinline[2][]
{%
  \bgroup%
    \lstset{#1}%
    \applyCurrentFontsize%
    \lstinline|#2|%
  \egroup%
}

\newcommand{\lstinlineKotlin}[1]{\scaledlstinline[language=Kotlin]{#1}}

\begin{document}
\title{Iterative and Scenario-based Requirements Specification in a System of Systems Context}
\titlerunning{Iterative and Scenario-based Requirements Specification}

\author{Carsten Wiecher\inst{1}\orcidID{0000-0002-3280-4471} \and
Joel Greenyer\inst{2}\orcidID{0000-0003-0347-0158} \and \\
Carsten Wolff\inst{1}\orcidID{0000-0003-3646-5240} 
\and
Harald Anacker\inst{3}
\and 
Roman Dumitrescu\inst{3}
}

\authorrunning{Wiecher et al.}


\institute{Dortmund University of Applied Sciences and Arts, 44139 Dortmund, Germany \\
\email{firstname.lastname@fh-dortmund.de}\and
FHDW Hannover, 30173 Hannover, Germany\\
\email{joel.greenyer@fhdw.de}
\and
Fraunhofer IEM, 33102 Paderborn, Germany\\
\email{firstname.lastname@iem.fraunhofer.de}
}

\maketitle              % typeset the header of the contribution
\begin{abstract}
[Context \& Motivation]
Due to the managerial, operational and evolutionary independence of constituent systems (CSs) in a System of Systems (SoS) context, top-down and linear requirements engineering (RE) approaches are insufficient. RE techniques for SoS must support iterating, changing, synchronizing, and communicating requirements across different abstraction and hierarchy levels as well as scopes of responsibility.
[Question/Problem] 
We address the challenge of SoS requirements specification, where requirements can describe the SoS behavior, but also the behavior of CSs that are developed independently. 
[Principal Ideas]
To support the requirements specification in an SoS environment, we propose a scenario-based and iterative specification technique. This allows requirements engineers to continuously model and jointly execute and test the system behavior for the SoS and the CS in order to detect contradictions in the requirement specifications at an early stage.  
[Contribution] 
In this paper, we describe an extension for the scenario-modeling language for Kotlin (SMLK) to continuously and formally model requirements on SoS and CS level. To support the iterative requirements specification and modeling we combine SMLK with agile development techniques. We demonstrate the applicability of our approach with the help of an example from the field of e-mobility.

\keywords{System of Systems Engineering  \and Requirements Analysis \and Requirements Specification \and Scenario-based Requirements Modeling}
\end{abstract}
\section{Introduction}
New methods and tools are needed to meet the challenges in the development of complex socio-technical systems, such as sustainable mobility solutions in metropolitan regions \cite{Ncube2018}. 
Systems of connected electrified vehicles can be characterised as a \textit{system of systems} (SoS), where the vehicle can be seen as a \textit{constituent system} (CS) that interacts with changing other CSs to provide an SoS functionality \cite{Hoehne2018}.

An interdisciplinary approach for the realization of these systems is \textit{system of systems engineering} (SoSE). 
The definition of stakeholder needs and required functionalities are key elements of SoSE \cite{INCOSE2015}; the precise specification of requirements is a basis for the system decomposition and implementation, or the selection of suitable CSs that form an SoS \cite{Odusd2008}. However, in SoSE, there are different requirements engineering (RE) challenges compared to RE in established systems engineering (SE) processes \cite{Nielsen2015}.%, \textcolor{limegreen}{that are targeting the development of complex monolithic systems as the previous generations }

According to Maier et al. \cite{Maier1996}, the operational, managerial, and evolutionary independence are the essential characteristics of an SoS. These characteristics have a significant influence on the applicability of existing RE techniques 
\cite{Ncube2011,Ncube2018,Nielsen2015}.
In contrast to monolithic systems, SoS consist of individual systems that can operate independently and perform a meaningful task, even when not part of an SoS. The development and operation of the CSs is managed independently, in different organizations with different development- and product life cycles. Also, requirements on the CS- and SoS level change frequently and independently, leading to an evolutionary development \cite{Nielsen2015,Maier1996}. 

Based on these SoS characteristics, Ncube and Lim \cite{Ncube2011} describe challenges for the SoS RE process: Due to the different systems in an SoS, requirements cover many different disciplines, can be contradictory, unknown or possibly not fully defined. These difficulties overlap with the fundamental problems in RE \cite{Fernandez2013}, but, according to Ncube and Lim \cite{Ncube2011}, requirements in an SoS additionally must be considered as requirements for the SoS, which describe the properties of the overall system, or requirements for a CS that describe capabilities of a single system. Since requirements on both levels can change continuously and independently, traditional, linear and top-down requirements specification and decomposition techniques can not be used \cite{Ncube2011,Ncube2018,Nielsen2015}. 

To address this problem we propose an iterative and scenario-based requirements specification technique.
Based on previous work \cite{Wiecher2020a,Wiecher2019,Wiecher2020} we integrate the Scenario Modeling Language for Kotlin (SMLK) with agile development techniques to support the requirements engineer in the continuous and iterative specification, formalization, and validation of requirements on different levels of abstraction. 

This paper makes the following two contributions:
First (1), we extend SMLK to enable requirements engineers to intuitively, but formally model the requirements on the SoS-level as well as the interaction between the CSs (CS-level). With these extensions, requirements can be specified and validated independently, which addresses the managerial and operational independence of systems. Nevertheless, both levels of abstraction are connected to allow for the joint execution and testing of the specified behavior on the SoS- and CS-level, in order to detect and resolve contradictions in the requirements on both these levels.   

Second (2), we propose a specification method where we combine behavior-driven development (BDD) and test-driven development (TDD) with the scenario-based modeling technique. This enables the iterative specification of system features and usage scenarios to document stakeholder expectations and generate tests steps, which subsequently drive the scenario-based modeling of the system specification.

While numerous approaches exist that suggest using formal scenario models to bridge the gap from informal requirements to the implementation of software-intensive systems \cite{Damas2006,Whittle2000,Harel2002d,Sutcliffe2003}, the particular contribution of this paper is the extension of sce\-nario-based modeling and programming techniques based on LSC Play-Out~\cite{Harel2002d} and behavioral programming (BP)~\cite{Harel2012} with BDD and TDD. Enabling this combination of agile development techniques with scenario-based requirements modeling addresses the \textit{coverage and sampling concerns} in scenario-based requirements engineering \cite{Sutcliffe2003}: by connecting features with tests (BDD), and tests with the scenario-based requirements model (TDD), we can ensure that every feature is modeled by an appropriate set of scenarios, and that these scenarios are validated by an appropriate set of tests.   

We asses the applicability with a proof-of-concept e-mobility application and provide a demonstration tool\footnote{https://bitbucket.org/crstnwchr/besos (includes the proof-of-concept example)}\footnote{https://bitbucket.org/jgreenyer/smlk/ (required to build the example project)} to enable others to use, evolve, and evaluate our approach.

Structure: We describe background in Sect.\,\ref{sec:background}, the scenario-based requirements specification method in Sect.\,\ref{sec:SosRequirementsSpecification}, and the proof-of-concept application in Sect.\,\ref{sec:proofOfConcept}. We report related work in Sect.\,\ref{sec:relatedWork} and conclude in Sect.\,\ref{sec:outlook}.

\section{Background}
\label{sec:background}
\subsection{System of Systems Engineering (SoSE)}
For the description of System of Systems (SoS) no generally valid definition yet exists \cite{Albers2018,Nielsen2015}. Hence, a distinction between complex monolithic systems and SoS is often made by the system characteristics. Therefore Maier describes five key characteristics of SoS \cite{Maier1996}: 
(1) \textit{Operational Independence}: Each system that is part of the SoS is independent and can perform a meaningful task, even if it is not integrated into the SoS.   
(2) \textit{Managerial Independence}: The individual systems are self-administered and individually managed. Consequently they collaborate with the other systems of the SoS, but they operate independently. 
(3) \textit{Geographic Distribution}: The individual systems of the SoS are distributed over large spatial distances, which means that the exchange of information between the individual systems is of primary importance for collaboration. 
(4) \textit{Evolutionary Development}: The objectives and functionality of an SoS can change constantly, as they can be added, modified or removed based on experience. Therefore an SoS never appears to be fully completed.
(5) \textit{Emergent Behavior}: By the collaboration of the individual systems, a synergism is achieved in which the SoS fulfils a purpose that cannot be achieved by or attributed to any of the individual systems.

These characteristics have a strong influence on the SoS development. To support a structured SoS development Dahmann et al. \cite{Dahmann2008} describe the differences between systems engineering (SE) and SoS engineering (SoSE). 
Accordingly, SE and SoSE both start with identifying and understanding user capability objectives in order to derive technical requirements for the system to be developed. In SE we subsequently continue with a top-down requirements decomposition and system design, with clear responsibilities in the management and engineering of the system  \cite{Gausemeier2002}. 
In SoSE, by contrast, the identified objectives and requirements serve as a basis for the development of new systems \emph{or} the integration of existing systems to build the SoS. Particularly the operational and managerial independence of individual systems is challenging: the existing systems may also fulfill other purposes that may conflict with the SoS objectives and those of its CS. Therefore it is important to understand how the individual systems behave and how this behavior contributes to the overall SoS behavior. 

\begin{figure}[h]
	\centering
	\includegraphics[width=0.5\linewidth]{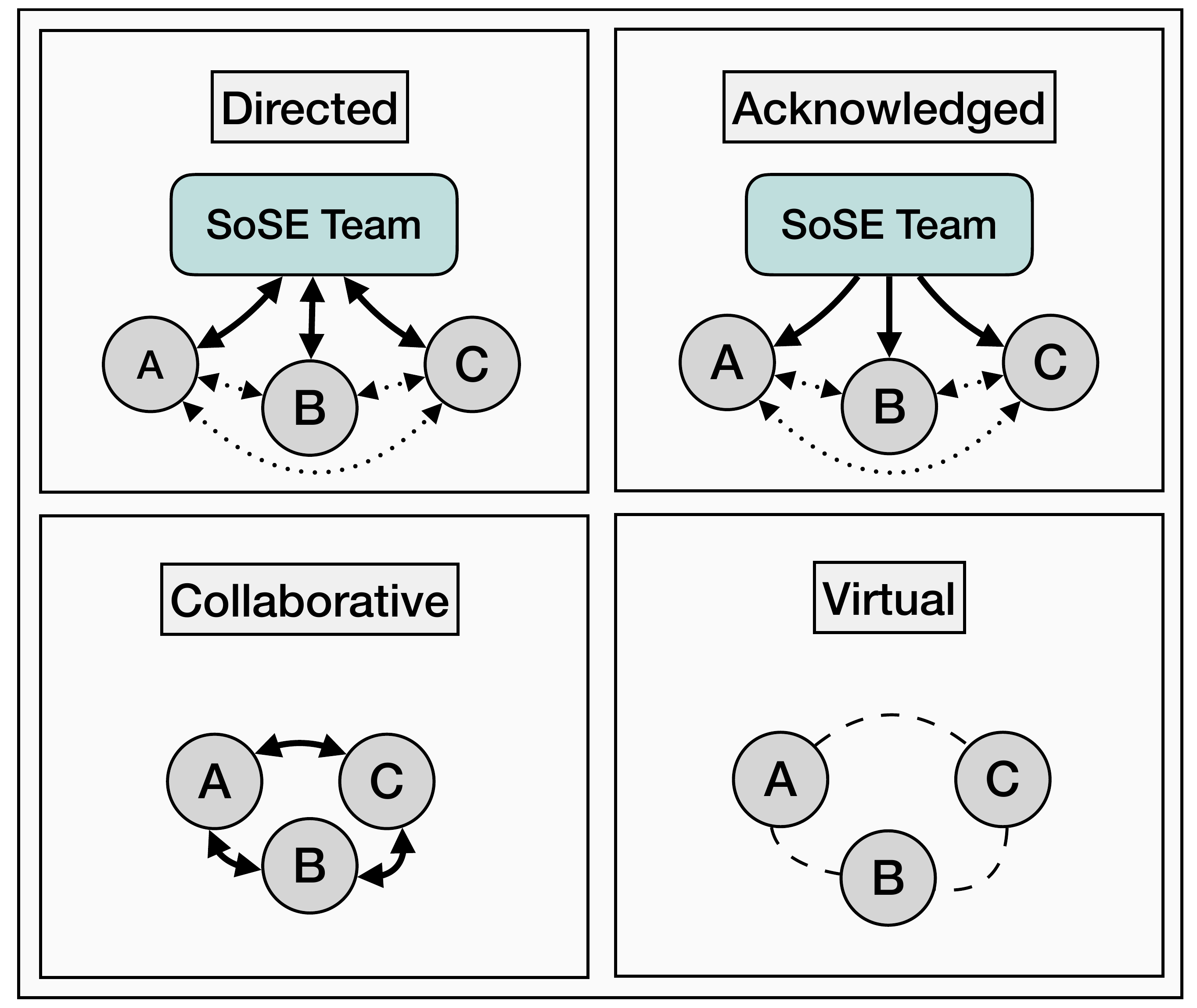}
	\caption{Different types of SoS \cite{Maier1996,Dahmann2008}}
	\label{fig:SoSTypes}
\end{figure}
When starting the SoS development it is important to categorize the SoS to be developed at an early stage because this has a significant influence on the RE approach that can be applied \cite{Nielsen2015,Dahmann2008,Ncube2018}. Fig.\,\ref{fig:SoSTypes} shows four different SoS types, initially introduced by Maier \cite{Maier1996} and extended by Dahmann and Baldwin \cite{Dahmann2008}:
A \emph{directed} SoS is designed for specific purposes. The individual systems have the ability to operate independently but are managed by a SoSE Team in a way that they fulfill a specific purpose. In an 
\emph{acknowledged} SoS the SoSE Team recognises and defines a common purpose and goal, but the CSs retain independent control and goals. The continuous and evolutionary development of the common purpose is based on collaboration between the SoS and the CSs. 
In a \emph{collaborative} SoS the individual systems are not bound to follow a central management, but voluntarily participate in a collaboration in order to achieve the SoS goal. 
A \emph{virtual} SoS has neither a leading control nor a common goal. This leads to a high degree of emergent behavior where the exact means and structures that produce the functionality of the system are difficult to recognize and distinguish \cite{Nielsen2015,Odusd2008}. 

This paper focuses on acknowledged SoS and we introduce an example next.

\subsection{Example of Application}
\label{sec:ExampleOfApplication}
To illustrate our approach, we introduce an e-mobility system of systems. In \cite{Kirpes2019} Kirpes et al. introduce an architecture model that provides an integrative view on former separated areas of electricity, individual mobility, and information and communication technologies to realize future e-mobility SoS.  

\begin{figure}[ht]
	\centering
 	\includegraphics[width=1.0\linewidth]{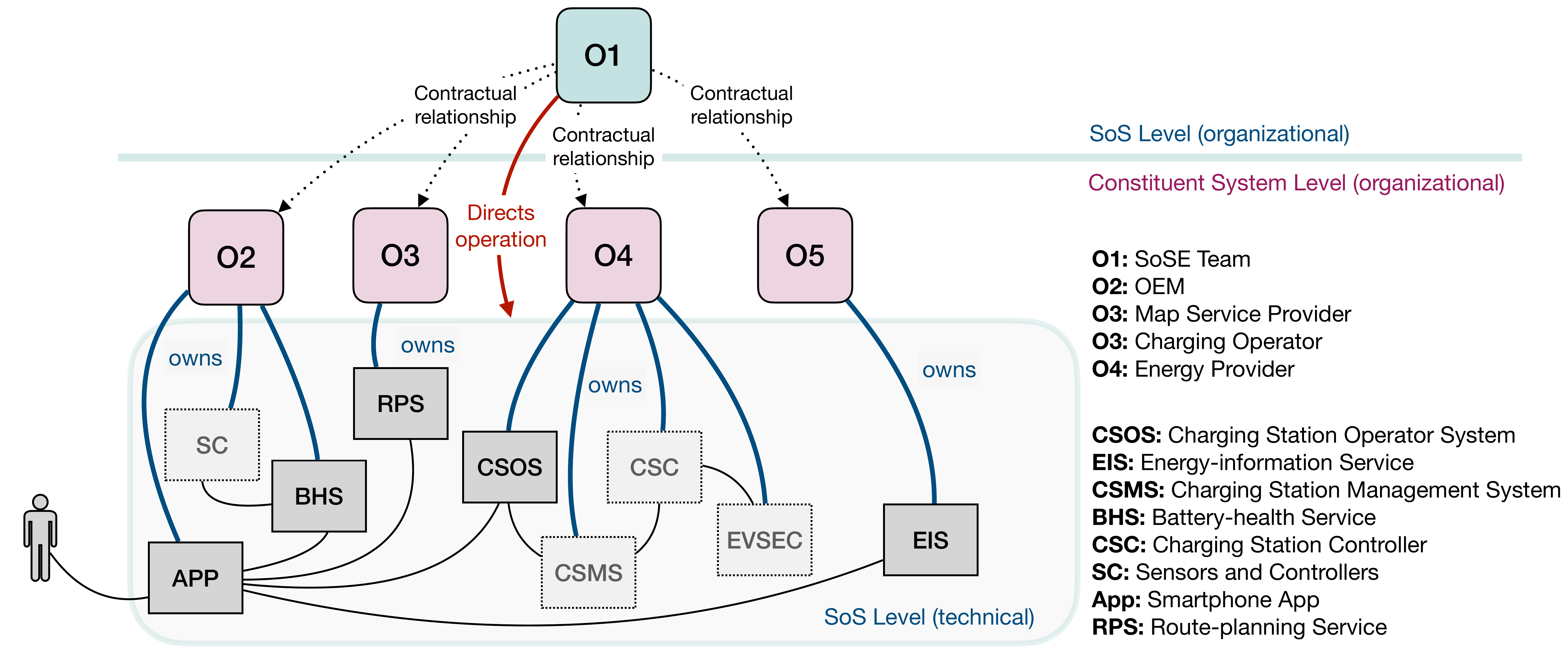}
	\caption{Smart charging as an acknowledged SoS. Based on \cite{Ncube2018} and \cite{Kirpes2019}.}
	\label{fig:SoSExample}
\end{figure}

Based on the example defined in \cite{Kirpes2019}, Fig.\,\ref{fig:SoSExample} shows an SoS user who is interacting with an e-mobility SoS. The main interest of the user is to improve the e-mobility experience and to reduce its costs. 
These user interests are targeted by a high-level use case that describes how to create an optimized travel plan. At the beginning the user enters travel preferences like start and destination into the smartphone app (APP). The APP then requests further data from other systems that are necessary for the calculation of an optimized route. 
For example, GPS data of possible routes are requested from a route-planning service (RPS). Usage data of available charging points along the routes are provided by a charging station operation service (CSOS). Recommendations for a battery-saving charging process are provided by a battery-health service (BHS). And information on current electricity prices in the region is provided by the energy-information service (EIS). 

These different systems, which are required to provide information to calculate an optimal route based on user preferences, are developed and managed by four different system owners (OEM, Map Service Provider, Charging Operator, and Energy Provider). We also see one SoSE Team which defines the overall SoS functionality, directs the operations, and has a contractual relationship with the owners of the CSs. According to \cite{Maier1996} and \cite{Dahmann2008} this example has the characteristics of an acknowledged SoS: we have recognised requirements, objectives and responsibilities on the SoS level and a contractual relationship between the SoSE Team and the individual constituent systems owner. However, the constituent systems keep their own management, funding and development approaches (cf. \cite{Ncube2018}).   

\subsection{Scenario Modeling Language for Kotlin (SMLK)}
\label{sec:SMLK}

SMLK is a Kotlin-based implementation of the Behavioral Programming (BP) paradigm \cite{Harel2012}. In BP, a program consists of a number of \textit{behavioral threads}, which we also call \textit{scenarios}. Scenarios are loosely coupled via shared events and can model individual behavioral aspects or functional requirements of a system. Scenarios can \textit{request} events that shall happen, be \textit{triggered by} or \textit{wait for} events requested by other scenarios, or (temporarily) \textit{forbid}/\textit{block} events. During execution, the scenarios are interwoven to yield a coherent system behavior that satisfies the requirements of all scenarios.

\begin{figure}[ht]
	\centering
	\includegraphics[width=1.0\linewidth]{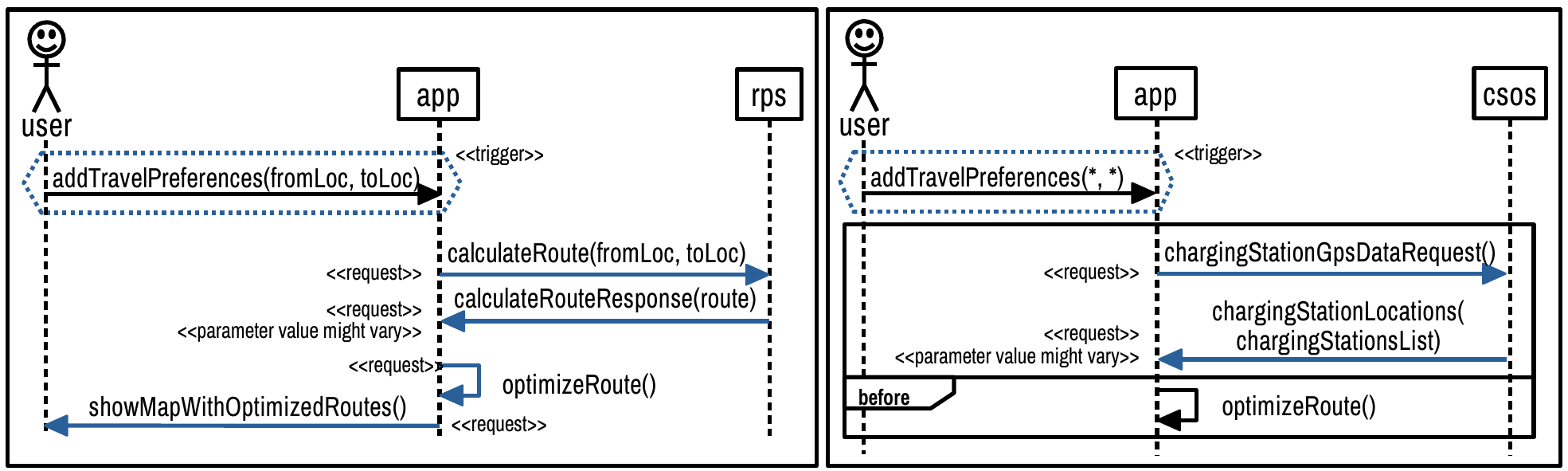}
	\caption{Graphical representation of the SMLK scenario in Listing~\ref{list:interSystemObjectEvent}}
	\label{fig:example-scenario-SD}
\end{figure}

Listing~\ref{list:interSystemObjectEvent} shows two SMLK scenarios that can be represented graphically as shown in Fig.\,\ref{fig:example-scenario-SD}.
Both scenarios are triggered by the event of a user entering the travel preferences in the app. This event is modeled as an interaction event of the object \lstinlineKotlin{user} sending the object \lstinlineKotlin{app} a message 
\lstinlineKotlin{addTravelPreferences}. In the first scenario, the parameters 
\lstinlineKotlin{fromLoc} and 
\lstinlineKotlin{toLoc} are variables bound to the parameter values carried by the triggering event when the scenario is triggered and initialized.
The SMLK code in the listing shows this binding of the parameter values explicitly (lines 2 and 3). The second scenario is triggered by the same event, but does not use the parameter values; the sequence diagram expresses this by using asterisks.

After the trigger event, the first scenario requests that the \lstinlineKotlin{app} sends the Route Planning Service (\lstinlineKotlin{rps})
a message to calculate the route between
\lstinlineKotlin{fromLoc} and 
\lstinlineKotlin{toLoc}, and then requests that the \lstinlineKotlin{rps} shall respond with a route. Then the \lstinlineKotlin{app} shall optimize the route and show it to the user.

The second scenario describes the interaction of the \lstinlineKotlin{app} and the Charging Station Operating System (\lstinlineKotlin{csos}). After the triggering event, the scenario requests that the \lstinlineKotlin{app} sends the \lstinlineKotlin{csos} a request to send GPS position data of available charging stations. The scenario then requests that the \lstinlineKotlin{csos} shall respond with such a list. This interaction must happen before the app optimizes the route, i.e., the event \lstinlineKotlin{app.optimizeRoute()} is blocked until the second scenario terminates; only then can be first scenario proceed.

In these example scenarios, the route details and charging location list contents are not relevant, so mock instances are created by helper functions. When at a later point the behavior is refined, these parameter values may be replaced by other values, e.g., a detailed and correct route may be calculated elsewhere. The scenario method \lstinlineKotlin{requestParamValuesMightVary} allows us to request events with supplied default parameter values, but it will accept also events sent between the same objects, and with the same signature, but with different parameter values.

\vspace{-0.1cm}
\begin{lstlisting}[caption=Example scenario from the e-mobility system specification,
	label=list:interSystemObjectEvent,
	style=KotlinStyle
	]
scenario(user sends (app receives App::addTravelPreferences)){
    val fromLoc = it.parameters[0] as String
    val toLoc = it.parameters[1] as String
    request(app sends rps.calculateRoute(fromLoc, toLoc))
    val route = createMockRoute()
    requestParamValuesMightVary(rps sends app.calculateRouteResponse(route)
    request(app.optimizeRoute())
    request(app sends user.showMapWithOptimizedRoute())
},
scenario(user sends (app receives App::addTravelPreferences)){
    scenario {
        request(app sends csos.chargingStationGpsDataRequest())
        val chargingStationsList = createMockChargingStationsList()
        requestParamValuesMightVary(csos sends app.considerChargingStationLocations(chargingStationsList))
    }.before(app.optimizeRoute())
}
\end{lstlisting}

\section{Scenario-based Requirements Specification in a System of Systems Context}
\label{sec:SosRequirementsSpecification}

To develop an SoS, usually existing systems are integrated by new systems to comprise a new SoS. While the new systems may be under a direct managerial and operational control, existing systems may be under the managerial and operational control of another organization.
Over time, systems that are under external control may change, which leads to the necessity to continuously (1) analyze how the changes in one system impact the SoS functionality, and (2) how other systems may have to be adapted to ensure that the SoS functionality can still be provided. This requires the SoSE team to continuously analyze, specify, and align requirements across different hierarchy levels. 

Our scenario-based requirements specification approach supports an iterative and integrated behavior modeling and analysis on the SoS and CS level. 

Based on the definitions in \cite{Harel2020} we introduce the term \emph{inter-system scenarios} to model the behavior on the SoS level and \emph{intra-system scenarios} to model the CS behavior. Also we show how both views can be integrated to allow for the joint execution and testing of the integrated SoS and CS behavior.

\subsection{Inter-System Scenarios}

The goal of modeling inter-system scenarios is to conceive an validate how SoS use cases can be realized by the interaction of users, existing systems, and new systems to be developed.
The inter-system scenario modeling process starts by defining the use cases, the structural SoS architecture, and then detailing and validating the use cases using scenarios and repeated simulation.

When modeling this behavior, certain assumptions are made about the behavior of the existing systems, possibly based on available documentation or communication with experts from the respective organizations.

Two exemplary inter-system scenarios are already introduced in Listing \ref{list:interSystemObjectEvent}, where we first modeled the interaction between the app, rps and the SoS user, and in the second scenario, between the user, app and csos. 
In this example, we see that we are able to model the interaction between selected systems, where new requirements can be considered by iteratively adding new scenarios to the \emph{SoS scenario specification}. By adding these inter-system scenarios the introduced modeling concepts allow to focus on a high level system interaction; 
Although we are able to partly ignore specification details (e.g. exact route information in Listing \ref{list:interSystemObjectEvent} line 14), we are able to execute and validate the interaction between the CS. This supports on the SoSE team to get a better understanding of the overall system behavior.

\subsection{Intra-System Scenarios}
\label{sec:intraSystem}

Once a satisfactory concept of the inter-system behavior is established, the inter-system specification must be supplemented and refined in two ways:
First (1), it is necessary to specify the behavior of the existing systems in more detail in order to validate whether the inter-system interaction behavior is indeed aligned with the behavior of the existing systems. 
Second (2), the behavior of the new systems to be developed must be detailed, possibly detailing their component structure and internal interactions, in order to provide a thorough basis for their development.

Our approach supports modeling the behavior on this more detailed hierarchy level with scenarios as well, and even to integrate their execution in order to simulate and validate behavioral requirements consistency across the different hierarchy levels.

\begin{figure}[ht]
	\centering
	\includegraphics[width=1.0\linewidth]{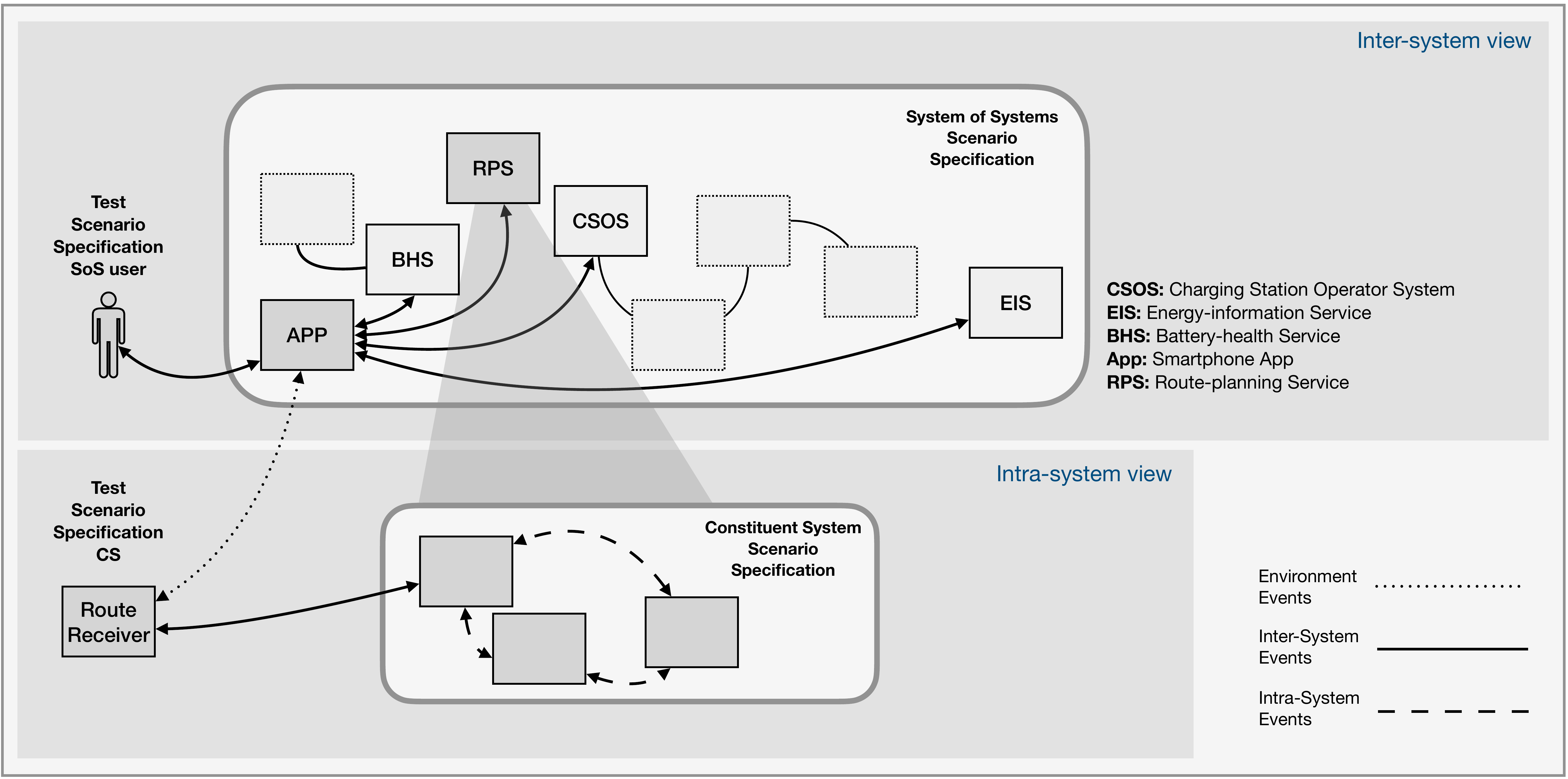}
	\caption{Inter- and intra-system view to continuously concretise requirements on CS level, while also considering the overall SoS behavior.}
	\label{fig:interactionSosCsLevel}
\end{figure}

To better distinguish between these two hierarchy levels, we distinguish the \textit{inter-system} level and \textit{intra-system} level as outlined in Fig.\,\ref{fig:interactionSosCsLevel}. 
The SoS scenario specification is located in the inter-system view, and individual CS scenario specifications are located in the intra-system view. When defining the internal behavior of a selected CS, we switch the perspective from the SoSE Team to a systems owner who is responsible for the development of a system. This can be e.g. the map service provider who is responsible for the development of the rps (see Fig.\,\ref{fig:SoSExample}). 

The intra-system scenarios are added to an individual CS scenario specification, with the goal to model requirements which are needed to build the CS and its subsystems. One example intra-system scenario is shown in Listing \ref{list:intraScenarioSpecification}.
\begin{lstlisting}[caption=CS scenario specification of the RPS,
	label=list:intraScenarioSpecification,
	style=KotlinStyle
	]
scenario(routeRequester sends(rps receives Rps::calculateRoute)){
    val fromLocString = it.parameters[0] as String
    val toLocString = it.parameters[1] as String
    request(rpsController sends gpsService.getLocations(fromLocString, toLocString))
    val fromLoc = getLocation(fromLocString)
    val toLoc = getLocation(toLocString)
    request(gpsService sends rpsController.locations(fromLoc, toLoc))
    request(rpsController sends routePlaner.calculateRoute(fromLoc, toLoc))
    val route = calculateRoute(fromLoc, toLoc)
    request(routePlaner sends rpsController.calculatedRoute(route))
    request(rpsController sends routeRequester.calculateRouteResponse(route))
}
\end{lstlisting}

The scenario specifies how the internal components of the rps (rpsController, gspService, and routePlanner) interact when receiving a request to calculate a route. Eventually (line 11), the calculated route will be returned to the requesting object.

When looking at the scenario in more detail, we see that the scenario is triggered when a \lstinlineKotlin{routeRequester} sends the \lstinlineKotlin{rps} the message \lstinlineKotlin{calculateRoute}.
This event is requested on the inter-system level, see the first scenario in Listing~\ref{list:interSystemObjectEvent} (line 4). 

One difference is, however, that in the intra-system scenario, we abstract from the app as being the source of the \lstinlineKotlin{calculateRoute} request (and the recipient of the route as a reposonse, see line 11). Instead, we assume that there is an abstract external route-requesting entity that requests a route to be calculated by the rps. We do this to separate the intra-system specification of a system from the particular SoS context defined on the inter-system level, as the system may also be used in other contexts.

The inter-system and intra-system level scenario execution can nevertheless be integrated, because the type of \lstinlineKotlin{routeRequester} is an interface that is also implemented by \lstinlineKotlin{app} (without showing the code in more detail for brevity). Hence it is possible that the event of the app requesting to calculate a route triggers the scenario shown here, and indeed the app would then receive the calculated route as a response.

The event parameters on the intra-system level may vary or be more detailed than the values assumed on the inter-system level where, for example, we used simple mock values (see Listing~\ref{list:interSystemObjectEvent}, lines 5 and 13). It is possible for intra-system scenarios to provide more detailed parameter values where the inter-system level scenarios request events by using the \lstinlineKotlin{requestParamValuesMightVary} command. (see Listing~\ref{list:interSystemObjectEvent} line 14).

\subsection{Specification Method}
\begin{figure}[ht]
	\centering
	\includegraphics[width=1.0\linewidth]{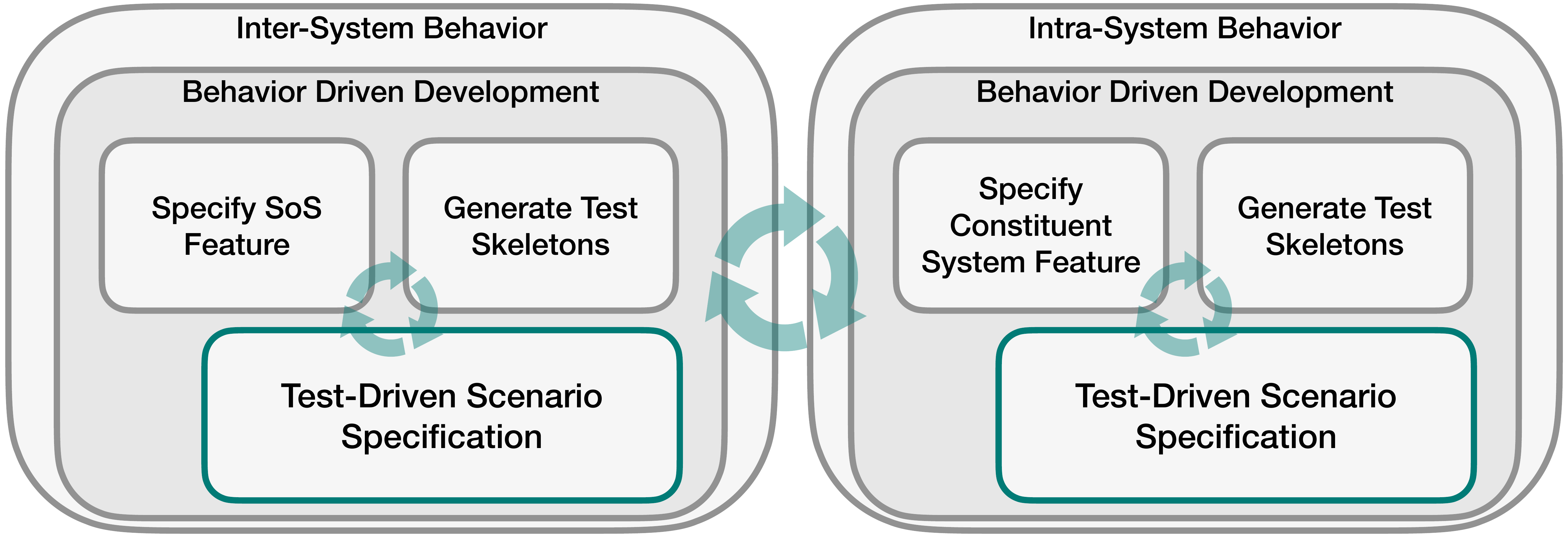}
	\caption{Continuous and iterative scenario specification}
	\label{fig:iterativeSpecification}
\end{figure}

To support the requirements engineer in modeling system requirements with SMLK, we propose an iterative method based on agile techniques. Fig.\,\ref{fig:iterativeSpecification} shows an overview of the single steps. We start with the specification of the inter-system behavior by applying the BDD approach. Here we first define the expected system behavior from the SoS user perspective.
Therefor we create a \emph{SoS feature specification} where each feature is defined by one or more \emph{usage scenarios} written in the gherkin syntax\footnote{https://cucumber.io/docs/gherkin/}. 
Listing~\ref{list:gherkinFeature1} shows a first feature specification that describes a user interaction with the app. On this hierarchy level, the SoS feature specification allows the SoSE team to define what is expected from the SoS and to document this expectations in a comprehensible form. 
\begin{lstlisting}[caption=Initial feature specification including a usage scenario to describe the user interaction with the SoS.,
	label=list:gherkinFeature1,
	style=GherkinStyle
	]
Feature: Retrieve travel preferences and display optimized route

    Scenario: Add travel preferences to the app
        When the SoS user adds travel preferences to the app
        Then the app displays a set of optimized routes
\end{lstlisting}
Based on this SoS feature specification we generate test skeletons as shown in Listing~\ref{list:testSkeletons}. 
These test skeletons are then used to drive the modeling of the inter-system behavior. To support a structured and iterative modeling of system requirements, we embed the Test-Driven Scenario Specification (TDSS) \cite{Wiecher2019} into the BDD approach. In this way, we combine the comprehensible specification of expected system behavior with the formal and scenario-based modeling of system requirements. 
\begin{lstlisting}[caption=Generated test steps.,
	label=list:testSkeletons,
	style=GherkinStyle
	]
When("^the EV user adds travel preferences to the App$") {
    //implement here     
}
Then("^the App displays a set of optimized routes$") {
    //implement here 
}
\end{lstlisting}

\begin{figure}[ht]
	\centering
	\includegraphics[width=1.0\linewidth]{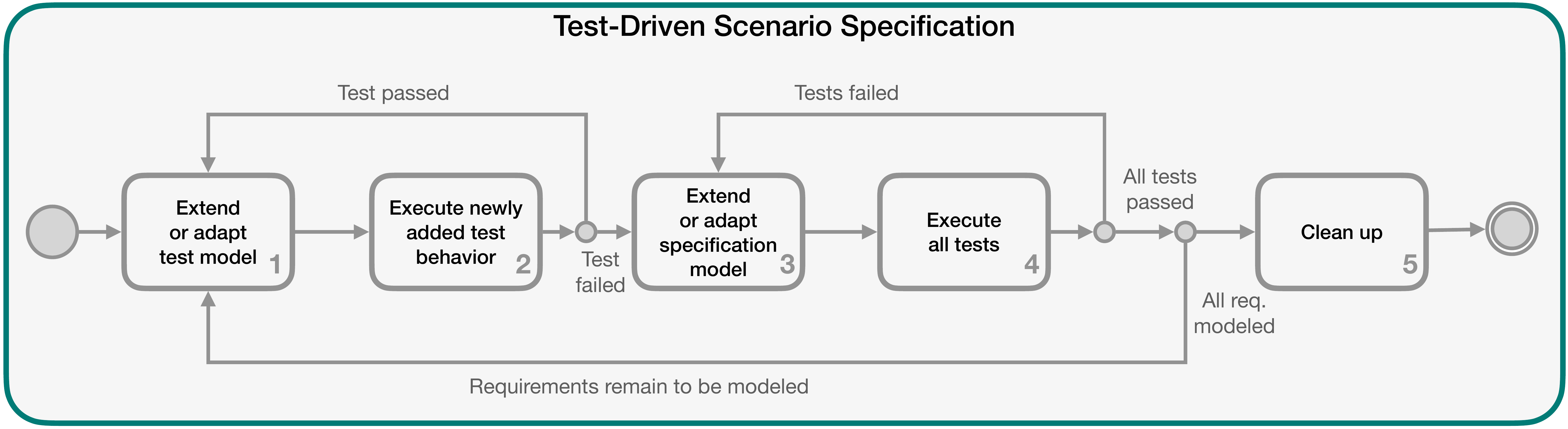}
	\caption{Test-Driven Scenario Specification (TDSS) \cite{Wiecher2019}}
	\label{fig:tdss}
\end{figure}

The TDSS approach includes the steps outlined in Fig.\,\ref{fig:tdss}. In the first step we extend the generated test skeletons (1). Here, we e.g. model that the user adds travel preferences to the app (Listing~\ref{list:testSteps} line 2) and eventually receives a map with optimized routes (line 5).
After we added these functions we execute the SoS feature specification (2) whereupon the single test steps and finally the events within the test steps are executed. At this point in time we did not model the inter-system behavior and consequently the test fails, because the app will not send the optimized route to the SoS user as expected in line 5.   
\begin{lstlisting}[caption=Generated test steps.,
	label=list:testSteps,
	style=GherkinStyle
	]
When("^the EV user adds travel preferences to the App$") {
    trigger(user sends app.addTravelPreferences("Dortmund", "Paderborn"))
}
Then("^the App displays a set of optimized routes$") {
    eventually(app sends user.mapWithOptimizedRoutes())
}
\end{lstlisting}
Therefore we extend our SoS scenario specification with the inter-system scenarios (3) which we already introduced in Listing \ref{list:interSystemObjectEvent}. We then run the test again to ensure that the modelled system requirements meet the expectations (4). If we have modeled additional tests in previous iterations, we now run them as well to ensure that there are no unexpected interactions between the individual tests and system requirements. If there are more requirements that need to be modeled, we perform further iterations. When all requirements on the SoS level known at this time have been modeled and tested, the SoS feature specification can be cleaned up. Afterwards the detailed specification of selected systems under development follows. 

This iterative approach supports the modeling of the interaction of all CSs within the SoS. 
%By using SMLK we are able to add new systems and events which are exchanged between these systems by adding new scenarios. 
In this way we are able to iteratively document the expectations from an SoS user perspective and model and test the interaction between the CSs. Thereby new systems and behavior can be added as needed to realize the expected behavior. When we have gone through several iterations, the SoSE team gets a better understanding of which systems are needed and what information these systems have to exchange with each other. Subsequently we can switch to the intra-system level and focus on the requirements specification for a selected CS within the SoS. Based on our example outlined in Fig.\,\ref{fig:SoSExample} we now switch from the SoSE team perspective to e.g. the perspective of the map service provider, who is responsible for the development of the rps. As shown in Fig.\,\ref{fig:iterativeSpecification} we execute the same specification method, but we create an independent \emph{CS feature specification}, generate independent test steps and create an CS scenario specification.  
This allows the independent specification and modeling of the requirements for the CS, which addresses the managerial, operational and evolutionary independence of systems in an SoS. In this way, system requirements can be specified without seeing the system in an SoS context. But, at the same time, both views can be integrated (as described in Sec. \ref{sec:intraSystem}), which allows the joint execution of the SoS behavior and the internal behavior of single already specified systems. 
In this way it's possible to detect contradictions between requirements on both levels. For example, if requirements have been specified at CS level that appear to have nothing to do with the SoS behavior but still influence the expected SoS behavior, the joint execution of the scenario specifications can be used to detect and resolve these dependencies. 

\section{Proof of Concept}
\label{sec:proofOfConcept}
To assess the applicability of our approach we integrated SMLK with the BDD tool Cucumber and executed the previously described specification method based on the example introduced in Section \ref{sec:ExampleOfApplication}. 

\begin{figure}[ht]
	\centering
	\includegraphics[width=1.0\linewidth]{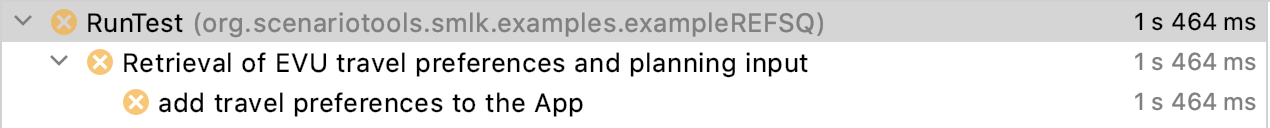}
	\caption{First TDSS run on SoS level}
	\label{fig:firstSosTest}
\end{figure}

\begin{figure}[ht]
	\centering
	\includegraphics[width=1.0\linewidth]{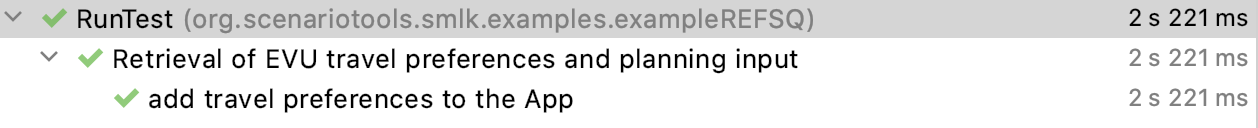}
	\caption{Execute tests after adapting the SoS scenario specification}
	\label{fig:secondSosTest}
\end{figure}

On SoS level we started with the feature specification as already shown in Listing~\ref{list:gherkinFeature1}. Subsequently we generated the test skeletons and added the SMLK events as shown in Listing \ref{list:testSteps}. 
Following the TDSS approach we executed the SoS feature specification (Step 1 in Fig.\,\ref{fig:tdss}) and got a failed test result as shown in Fig.\,\ref{fig:firstSosTest}. 
Subsequently we extended the SoS scenario specification as shown in Listing \ref{list:interSystemObjectEvent} to specify the SoS behavior. After we added these scenarios we executed the test again and finally received the expected event, resulting in a positive test result as shown in Fig.\,\ref{fig:secondSosTest}.

After we successfully defined a first interaction on  inter-system level, we switched to the intra-system level and added a CS scenario specification to model the internal behavior of the rps as shown in Listing \ref{list:intraScenarioSpecification}. 

Now we executed the same SoS feature again resulting in a negative result, because the rps internal behavior was not yet specified and hence the CS scenario program didn't send the \lstinlineKotlin{calculateRouteResponse(route)} message to the app.  

To fix this we executed the TDSS process within the intra-system view, based on the CS feature specification shown in Listing \ref{list:CSfeature}.
\begin{lstlisting}[caption=Feature on CS level,
	label=list:CSfeature,
	style=GherkinStyle
	]
Feature: Calculate route - RPS

    @RpsSystem
    Scenario: Calculate route based on user travel preferences 
    When the app sends travel preferences to the rps
    Then the rps responds route information including gps data 
\end{lstlisting}

Finally we got passed test results again, but now we also considered the rps internal behavior specification. And, by using \emph{tags} within the different feature specifications (e.g. \lstinlineKotlin{@RpsSystem}) and by applying the concepts described in Sect. \ref{sec:intraSystem}, we were not only able to validate the integrated SoS and CS behavior, but we also could independently test the requirements of single CS. 

To allow others to use, validate and evolve our approach, we describe the architecture and functional principles of the developed tool in \cite{Wiecher2021a} as a companion to this paper. Here, we also describe the method we outline in Fig.~\,\ref{fig:iterativeSpecification} in more detail. And we provide information about the necessary resources\footnote{https://bitbucket.org/crstnwchr/besos/}\footnote{https://bitbucket.org/jgreenyer/smlk/}\footnote{https://cucumber.io}\footnote{https://www.jetbrains.com/idea/} to build and execute the example we use in this paper.

\section{Related Work}
\label{sec:relatedWork}
In this paper we use SMLK, which was extended to support an iterative and continuous modeling of system behavior in an SoS context. This modeling language is based on Live Sequence Charts (LSCs) \cite{Damm2001}. A recent LSC variant are Modal Sequence Diagrams (MSDs) \cite{Harel2006}. By modeling behavioral requirements with the help of MSDs, different works argue that this formal requirements modeling can increase the requirements quality (e.g. \cite{Holtmann2014}, \cite{Fockel2016}),
but these approaches are based on traditional SE and do not consider the SoS characteristics and their impact on the requirements specification. 

Harel et al. describe an extension to behavioral programming that allows the integration of behavioral programs that operate on different hierarchy levels and time scales~\cite{Harel2011}. Indeed, we also use this approach to integrate different SMLK scenario programs that execute the behavior on the inter- and intra-system level.

Simulation-based analysis and design is commonplace in cyber-physical systems of systems, e.g. using actor-oriented frameworks or co-simulation \cite{Fitzgerald2014,Lee2015}. We aim to provide similar means for the thorough specification and analysis of \textit{requirements} of systems of systems. To the best of our knowledge, this is a new approach.

Other works address model-based RE in the SoS context. Holt et al. describes an ontology for model-based SoS requirements engineering \cite{Holt2012}. Albers et al. show how SoS requirements can be specified based on use-cases and sequence diagrams within SysML \cite{Albers2015}. However, an early, iterative and formal specification of requirements, with the goal to execute and test these requirements specifications is not considered in these approaches.  

\section{Summary and Outlook}
\label{sec:outlook}
In this paper, we propose a technology to continuously model behavior requirements in an SoS context. Our approach supports requirements engineers in the iterative specification, modeling and testing of requirements. With the use of SMLK, the system behavior can be modeled textually through scenarios. This scenario-based modeling is close to how engineers communicate system behavior and hence enables a feasible formalization of requirements. To further support and structure the formalization process, we integrated SMLK with agile techniques and appropriate tooling. This fosters the iterative formalization, and by testing the formalized requirements specifications, we get early feedback about the expected system behavior and possible contradictions in requirements. Due to the proposed coupling of inter- and intra- system scenarios, we are also able to execute and test the system behavior on different hierarchy levels. And by integrating the BDD tool cucumber, we are able to specify the expected system behavior with the help of features and usage scenarios written in natural language, which supports the communication of expected system behavior in a multi-disciplinary development team.

For future work, we plan to integrate our previous work \cite{Wiecher2019} and the modeling concepts shown in this paper with an automated test case creation proposed in \cite{Fischbach2020} to further reduce the modeling effort. Also, as already started in previous work \cite{Wiecher2020}, we plan to integrate the results of this paper in an automotive development process and validate the applicability within an ongoing research project. As shown in \cite{Wiecher2019}, we are able to find contradictions in automotive requirements specifications, but the open questions are if the approach is scalable and whether the effort for the requirements modeling is justified.

Another possible direction for future work is focusing on stakeholder needs in a SoS context. In this paper we already integrated the BDD approach to validate requirements and align stakeholder expectations. This could be done more systematically by integrating goal modeling approaches \cite{Aydemir2018}.

\bibliographystyle{splncs04}

\bibliography{sample-base}

\end{document}